\def\finalpaper{1} 

\def\replaceTAAL{1}

\documentclass[conference]{IEEEtran}
\usepackage{pgf-pie}  
\usepackage{dblfloatfix}
\usepackage{balance}
\usepackage{tikz}
\usepackage{amsmath}
\usepackage{amsfonts}
\usepackage[ruled,linesnumbered]{algorithm2e}
\usepackage{filecontents}
\usepackage{cuted,adjustbox}
\usepackage{multirow,multicol}
\usepackage{booktabs,colortbl}
\usepackage{pgfplots} 
\usetikzlibrary{patterns}
\usepackage[labelformat=simple]{subcaption}
\renewcommand\thesubfigure{(\alph{subfigure})}
\usepackage{rotating}
\usepackage{xcolor}
\definecolor{rwth_blue}{RGB}{0,84,159}
\definecolor{rwth_green}{RGB}{87,171,39}
\definecolor{rwth_red}{RGB}{204,7,30}
\usepackage{circledsteps}
\usepackage{xspace}
\usepackage{tcolorbox}
\usepackage{subcaption}
    \newcommand{\rulesep}{\unskip\ \vrule\ }
\tcbuselibrary{listings}
\newcommand\myCircled[2][]{\ifmmode
\Circled[fill color=black,inner color=white,#1]{\mathsf{#2}}
\else
\Circled[fill color=black,inner color=white,#1]{\sffamily#2}
\fi
}
\newtcblisting{foo}{
  listing only,
  nobeforeafter,
  after={\xspace},
  hbox,
  tcbox raise base,
  fontupper=\ttfamily,
  colback=lightgray,
  colframe=lightgray,
  size=fbox
  }
\usepackage{lipsum}

\usepackage{listings}
\definecolor{mGreen}{rgb}{0,0.6,0}
\definecolor{mGray}{rgb}{0.5,0.5,0.5}
\definecolor{mPurple}{rgb}{0.58,0,0.82}
\definecolor{backgroundColour}{rgb}{0.95,0.95,0.92}
\lstdefinestyle{CStyle}{
    backgroundcolor=\color{backgroundColour},   
    commentstyle=\color{mGreen},
    keywordstyle=\color{magenta},
    numberstyle=\tiny\color{mGray},
    stringstyle=\color{mPurple},
    basicstyle=\footnotesize,
    breakatwhitespace=false,         
    breaklines=true,                 
    captionpos=b,                    
    keepspaces=true,                 
    numbers=left,                    
    numbersep=5pt,                  
    showspaces=false,                
    showstringspaces=false,
    showtabs=false,                  
    tabsize=2,
    language=C
}

\usepackage{listings}

\usepackage{background}
\usepackage[usestackEOL]{stackengine}
\setstackgap{L}{\normalbaselineskip}
\SetBgContents{\color{blue}{\tiny \Longstack{PREPRINT - accepted at IEEE ISQED 2025.}}}
\SetBgPosition{4.5cm,1cm}
\SetBgOpacity{1.0}
\SetBgAngle{0}
\SetBgScale{1.8}

\usepackage{array}
\newcolumntype{x}[1]{>{\centering\arraybackslash\hspace{0pt}}p{#1}}

\AtBeginDocument{%
  \providecommand\BibTeX{{%
    \normalfont B\kern-0.5em{\scshape i\kern-0.25em b}\kern-0.8em\TeX}}}

\begin{document}
\bstctlcite{IEEEexample:BSTcontrol}


\title{The Impact of Logic Locking on Confidentiality: An Automated Evaluation\vspace{-0.3cm}}

\if\finalpaper1
\author{Lennart M. Reimann\IEEEauthorrefmark{1}, Evgenii Rezunov\IEEEauthorrefmark{1}, Dominik Germek\IEEEauthorrefmark{2}, Luca Collini\IEEEauthorrefmark{3}, \\ Christian Pilato \IEEEauthorrefmark{4}, Ramesh Karri \IEEEauthorrefmark{3}, and Rainer Leupers\IEEEauthorrefmark{1}\\
\IEEEauthorrefmark{1}RWTH Aachen University, Germany, 
\{lennart.reimann, rezunov, leupers\}@ice.rwth-aachen.de\\
\IEEEauthorrefmark{2}Corporate Research, Robert Bosch GmbH, Germany, dominik.germek@de.bosch.com\\
\IEEEauthorrefmark{3}NYU Tandon School of Engineering, USA, \{lc4976, rkarri\}@nyu.edu, \\\IEEEauthorrefmark{4}Politecnico di Milano, Italy, christian.pilato@polimi.it \\
\vspace{-1.3cm}
}

\else
\author{Anonymous Authors\\Affiliation\\Affiliation\vspace{-0.3cm}}
\fi

\maketitle

\begin{abstract}

Logic locking secures hardware designs in untrusted foundries by incorporating key-driven gates to obscure the original blueprint. While this method safeguards the integrated circuit from malicious alterations during fabrication, its influence on data confidentiality during runtime has been ignored.
In this study, we employ path sensitization to formally examine the impact of logic locking on confidentiality. By applying three representative logic locking mechanisms on open-source cryptographic benchmarks, we utilize an automatic test pattern generation framework to evaluate the effect of locking on cryptographic encryption keys and sensitive data signals. Our analysis reveals that logic locking can inadvertently cause sensitive data leakage when incorrect logic locking keys are used.
We show that a single malicious logic locking key can expose over 70\% of an encryption key. If an adversary gains control over other inputs, the entire encryption key can be compromised. This research uncovers a significant security vulnerability in logic locking and emphasizes the need for comprehensive security assessments that extend beyond key-recovery attacks.

\end{abstract}
\section{Introduction}

Modern Integrated Circuit (IC) supply chains rely on third-party design houses and foundries, which expose hardware design descriptions to external parties. Thus, one needs to prevent reverse engineering and malicious modifications by rogue entities, in a cost-effective way.  
While logic locking was initially conceived to safeguard ICs within the hardware supply chain against Intellectual Property (IP) piracy, subsequent research has explored its potential in thwarting reverse engineering and preventing malicious modifications to ICs with notable success~\cite{mentor, disquisition}. Logic-locking techniques address these threats, with the first commercially available logic-locked RISC-V processor, the ``Made in Germany RISC-V (MiG-V),'' demonstrating its applicability in an industrial setting~\cite{mig_v, mig-v2}.

\begin{figure}[t!]
    \centering
    \begin{subfigure}[b]{0.3\columnwidth}
         \centering
    \includegraphics[origin=c, width=0.84\columnwidth]{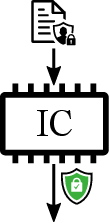 }
    \caption{not logic locked}
    \label{fig:not_locked}
    \end{subfigure}
    \rulesep
    \begin{subfigure}[b]{0.32\columnwidth}
         \centering
    \includegraphics[origin=c, width=0.84\columnwidth]{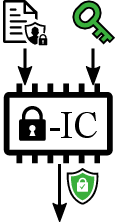 }
    \caption{locked, correct key}
    \label{fig:locked_correct_key}
    \end{subfigure}
    \rulesep
    \begin{subfigure}[b]{0.32\columnwidth}
         \centering
    \includegraphics[origin=c, width=0.8\columnwidth]{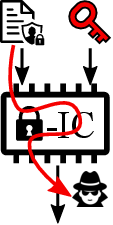 }
    \caption{locked, malign key}
    \label{fig:locked_malign_key}
    \end{subfigure}
    \caption{A logic-locked IC has the same functionality as its not-locked version (a) using the correct logic-locking key (b). The same functionality connotes the absence of direct data leakages. Misusing the logic-locking hardware with a malign key can cause sensitive data leakages (c). \label{fig:motivation}}
        \vspace{-0.4cm}

\end{figure}

The core principle of logic locking is to make the hardware design's functionality dependent on a secret logic locking key. Additional hardware, such as adders, XOR gates, or multiplexers, is incorporated into the IP, with the aim of distorting the design's functionality when applying the wrong logic locking key. The design is forwarded in the supply chain without the key. As the key is concealed from the untrusted parties, the IP's behavior cannot be easily derived from the hardware description, preventing the incorporation of \textcolor{red}{}malicious modifications within obscured segments of the hardware's functionality. 



However, as logic locking introduces additional hardware adaptations, the \textcolor{red}{}previously enforced properties can be endangered. By applying an ``incorrect'' logic locking key, the chip does not function as intended. Depending on the mechanics of the logic locking scheme, new signal paths or operations are added into the circuit. Thus, applying the ``incorrect'' key can activate undesired behavior, or \textit{introduce inadmissible data leakage paths} as depicted in Fig.~\ref{fig:motivation}. These additional paths can impose a major security vulnerability.
In a recent investigation, a manual security inspection uncovered exploitable vulnerabilities in the MiG-V's logic locking hardware that lead to sensitive encryption key leakage~\cite{mig_v_cracked}. These findings indicate that logic locking can create unintentional attack paths on sensitive components within hardware design.
To advance beyond manual inspection methods, we develop an automated approach to analyze how logic locking schemes affect information flow in hardware designs. We conduct this investigation using cryptographic circuits as benchmarks, given their fundamental role in protecting sensitive data.
We determine whether specific input sequences could leak the encryption key to the primary outputs of the design\footnote{The encryption key is used for data encryption/decryption and is distinct from the logic-locking key.}. This analysis is conducted before and after applying logic locking to the benchmarks, with the aim of \textit{determining if ``incorrect'' logic locking keys could be exploited to disclose sensitive data, e.g., cryptographic keys.} The key contributions of this work are threefold:
\begin{itemize}
    \item An automated evaluation of the impact of three \textit{representative} logic locking schemes on the confidentiality property of encryption keys in five cryptographic benchmarks. 
    \item An analysis of the impact of key length and algorithm choice on the level of threat exposure.
    \item A discussion on the hardware architecture's role in susceptibility to logic locking flaws.
\end{itemize}
\section{Background}
The following describes the logic locking schemes, the path sensitization method to find leakages, and the attack model.
\subsection{Logic Locking}\label{sec-ll-desc}
\begin{figure}[t!]
    \centering
    \includegraphics[width=\columnwidth]{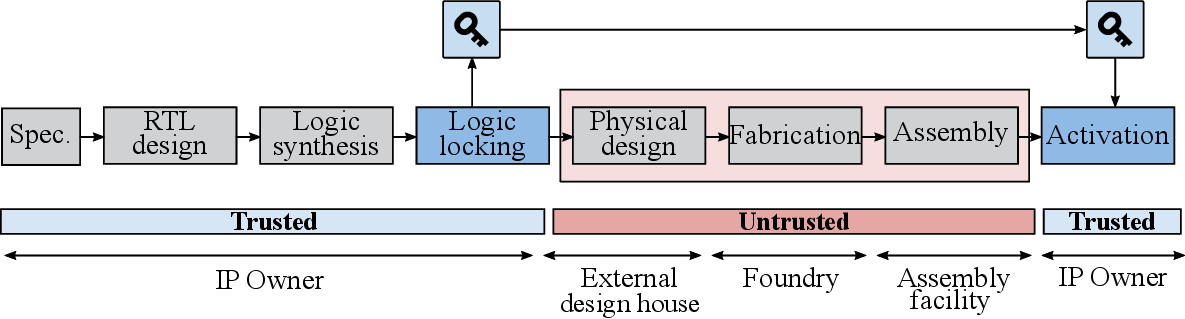 }
    \caption{Use of logic locking to secure the supply chain.\label{fig:supply_chain}}
    \vspace{-0.4cm}

\end{figure}

\begin{figure*}[t]
\begin{subfigure}[b]{\textwidth}
\renewcommand\thesubfigure{(\roman{subfigure})}

        \centering
        \begin{subfigure}{0.265\textwidth}
        \centering
             \includegraphics[width=0.95\textwidth]{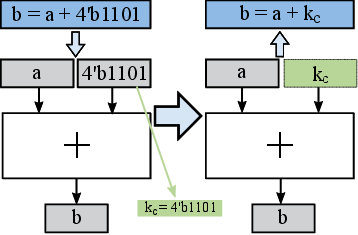 }
             \caption{Constant obfuscation\label{fig:assure_constant}}
        \end{subfigure}
         \rulesep
        \begin{subfigure}{0.34\textwidth}
        \centering
             \includegraphics[width=0.95\textwidth]{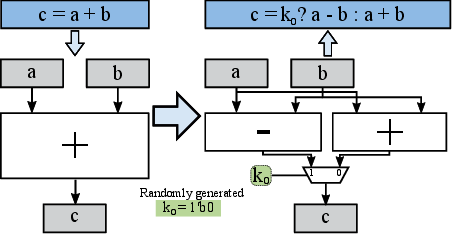 }
             \caption{Operation obfuscation\label{fig:assure_operation}}
        \end{subfigure}
        \rulesep
        \begin{subfigure}{0.365\textwidth}
        \centering
             \includegraphics[width=0.95\textwidth]{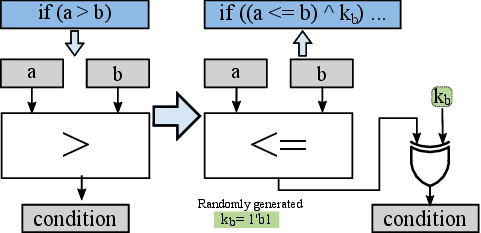 }
             \caption{Branch obfuscation\label{fig:assure_branch}}
        \end{subfigure}
        \setcounter{subfigure}{0}
    \renewcommand\thesubfigure{(\alph{subfigure})}
    \vspace{-0.4cm}
    \caption{ASSURE~\cite{assure}\label{fig:assure}}
    \vspace{0.2cm}
\end{subfigure}

    \centering
    \begin{subfigure}[b]{0.4\textwidth}
         \centering
    \includegraphics[origin=c, width=0.95\textwidth]{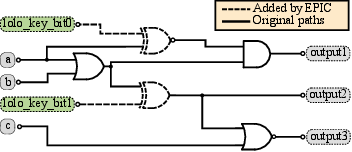 }
    \caption{EPIC~\cite{epic}}
    \label{fig:epic}
    \end{subfigure}
    \rulesep
    \begin{subfigure}[b]{0.4\textwidth}
         \centering
    \includegraphics[origin=c, width=0.95\textwidth]{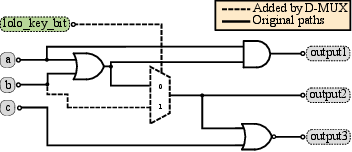 }
    \caption{D-MUX~\cite{dmux2022}}
    \label{fig:dmux}
    \end{subfigure}
    \caption{The three logic locking algorithms introducing additional logic: EPIC (adds XOR and XNOR gates), D-MUX (adds multiplexer), and ASSURE (adds logic on RTL level, such as additional ports, logic, and arithmetic operations and XOR gates).\label{fig:lolo_schemes}}
        \vspace{-0.4cm}

\end{figure*}

Logic Locking (LL) allows obfuscating a hardware design's functionality and structure, thus protecting the design from malicious manipulations (DoS hardware Trojans) while being processed in an untrusted external foundry~\cite{evolution, provably_lolo}. LL inserts additional key-controlled logic that binds the design's correct functionality to a secret key, which is only known to the legitimate IP owner. Locking is performed on the design before it reaches an external design house or foundry, as depicted in Fig.~\ref{fig:supply_chain}. The additional logic fuses with the existing design structure during logic synthesis, resulting in structural obfuscation. The LL key is embedded into the final chip after fabrication. The security of LL w.r.t. protecting against hardware manipulations is based on the assumption that a malicious entity must first find the correct activation key to reverse engineer, understand, and finally manipulate the design.
LL can be deployed at different design levels, including Register-Transfer Level (RTL) and gate level. 
In the following, we give more details on three different \textit{representative}  schemes: ASSURE~\cite{assure} (Fig.~\ref{fig:assure}), EPIC~\cite{epic} (Fig.~\ref{fig:epic}), and D-MUX~\cite{dmux2022} (Fig.~\ref{fig:dmux}). EPIC and D-MUX are representatives of two important classes of LL techniques on the gate level. ASSURE embodies the concepts of RTL locking. 

\paragraph{\textbf{RTL Locking (ASSURE)}}
Fig.~\ref{fig:assure} illustrates how ASSURE~\cite{assure} works. This LL scheme offers several modes to conceal the hardware design's functionality. To lock the RTL code, ASSURE employs a key that obfuscates operations, conditions, and constants. The process is as follows:
\begin{itemize}
    \item Constant Locking (Fig. 3(a)(i))  substitutes constants with corresponding key bits. For instance, the expression $b = a + 4'b1101$ is locked as $b = a + k_c$, where $k_c$ represents the 4-bit constant ($4'b1101$) stored within the locking key.
    \item Operation Locking (Fig. 3(a)(ii))  integrates a multiplexer to choose between the correct operation and a dummy operation, depending on a key bit. For instance, the expression $c = a \,+\, b$ is locked as:
\begin{lstlisting}[style=CStyle,numbers=none,mathescape=true]
        c = $k_o$ ? (a + b) : (a - b),\end{lstlisting}or
    \begin{lstlisting}[style=CStyle,numbers=none,mathescape=true]
        c = $k_o$ ? (a - b) : (a + b),\end{lstlisting}
    depending on the value of $k_o$.
    \item Branch Locking (Fig. 3(a)(iii)) modifies the condition by XORing it with a key bit. For example, the condition $a > b$ is locked as either $(a > b)^\wedge k_b$ or $(a <= b)^\wedge k_b$, based on value of $k_b$.

\end{itemize}
The locking key comprises two parts: one part is generated randomly and used for locking control branches and operations, while the other part contains constants extracted from the design. An input port is introduced to apply the locking keys after IC fabrication. A locking point refers to a semantic element, such as a constant, a branch, or an operation, which can be secured using these techniques. 
Securing a design at the RTL offers a compelling balance between protection and implementation. At this stage, the majority of semantic information, such as constants, operations, and control flows, remains available, allowing the obfuscation before information gets lost through synthesis optimizations. 

\paragraph{\textbf{XOR/XNOR Locking (EPIC)}}
A large branch of LL schemes is based on the insertion of key-controlled XOR/XNOR gates into the design. An example of this mechanism is presented in Fig.~\ref{fig:epic}. Here, the inserted XOR and XNOR gates are controlled via the key inputs $lolo\_key\_bit0$ and $lolo\_key\_bit1$. For $lolo\_key\_bit1=0$, the second input of the XOR gate is buffered to its output, thus preserving the original design's functionality. If $k_{0}=1$, the second input is inverted, resulting in erroneous behavior. The implemented XNOR key gate, introduces a locking mechanism in a similar fashion, except that the second input value is preserved for $lolo\_key\_bit1=1$. 
This fundamental mechanism has been integrated into various XOR/XNOR-based LL schemes~\cite{10.1145/2228360.2228377,7362173,9214869}. They differ in the specifics of the insertion strategy of the key-controlled gates.
As a random insertion represents a \textit{superset} of all strategies, further evaluations in this work are based on the EPIC scheme~\cite{epic}. 

\paragraph{\textbf{MUX-based Locking (D-MUX)}} The security of XOR/XNOR-based LL is built on top of the assumption that correlating the gate type (XOR/XNOR) with the correct key (0/1) is not possible. However, recent Machine Learning (ML)-based attacks have shown that this assumption is not valid~\cite{SAIL2018, sisejkovicSnapShot2021, 9539868}, since a structural analysis allows an educated guess about the correct key. To overcome this issue, Multiplexer (MUX)-based locking was introduced in the form of the deceptive MUX-based LL (D-MUX) scheme~\cite{dmux2022}. D-MUX inserts key-controlled MUX blocks thereby creating additional combinational paths within a design, as shown in Fig.~\ref{fig:dmux}. Hereby, the selection of the paths should avoid any form of logic-locking key-related information leakage that might be exploited by an ML model. 


The three LL algorithms, EPIC, ASSURE, and D-MUX, encompass a wide range of strategies in the field of LL. These algorithms operate on distinct abstraction layers and integrate various forms of logic into the original hardware. Consequently, this selection of schemes provides us with valuable insights into the potential threats to the confidentiality of sensitive signals posed by LL. 

\subsection{Path Sensitization}
In this work, we utilize path sensitization~\cite{path_sensitization} to determine input patterns that can forward the sensitive data (e.g., encryption keys) to the primary outputs of the hardware, allowing adversaries to extract the secret. Hereafter, Automatic Test Pattern Generation (ATPG)~\cite{testmax} is employed to conduct the required path sensitization.  

\begin{figure}[t!]
    \centering
    \includegraphics[width=\columnwidth]{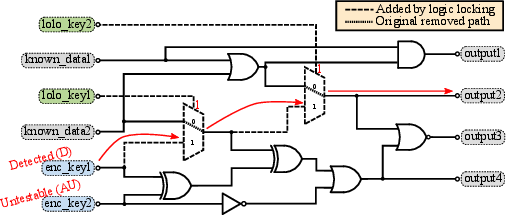 }
    \caption{Path sensitization is applied to retrieve the encryption key bits from the circuit. The analysis shows a detection for bit 1 (enc\_key1), by applying the logic locking key ``11''. No input combination of the known inputs and logic locking key bits can forward enc\_key2 to an output. A logic locking key of ``00'' restores the original functionality. The leakage of enc\_key1 occurs via paths introduced by logic locking.\label{fig:path_sensitization}}
    \vspace{-0.1cm}

\end{figure}
The ATPG framework determines the input sequence that establishes a path from the gate to an output, as depicted in Fig.~\ref{fig:path_sensitization}. As we would like to learn whether sensitive data can be leaked, we apply the framework to gather information about leakages, by marking the sensitive signal for the ATPG framework. If such an input sequence exists, the datum can be leaked and is labelled as \textit{detected (D)}.

If no input sequence exists that can activate a signal path between the sensitive bit and the output bit, the bit is labeled as \textit{secure (S)}. In our work, secure signal bits represent sensitive data that cannot be fully leaked to an adversary. For an encryption benchmark, every encryption key bit must be \textit{S}. Otherwise, an attacker could easily access the key. 

Consider the scenario illustrated in Fig.~\ref{fig:path_sensitization}. Here, the multiplexers added by a logic-locking algorithm can forward one of the encryption key bits. However, the second encryption key bit ($enc\_key2$) remains \textit{S}. 
In contrast, in the original hardware design, both encryption key bits remain untestable and therefore secure.
If the complexity of the design forbids the framework to determine whether the bit is \textit{D} or \textit{S} within the time limit, the bit is labeled as \textit{not detected (ND)}.

\subsection{Attack Model}
\if\replaceTAAL0

\begin{figure}[t!]
    \centering
    \includegraphics[width=\columnwidth]{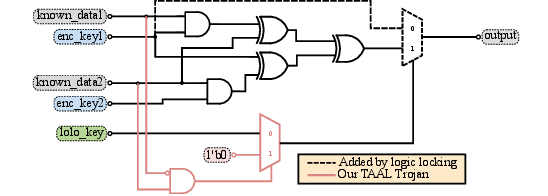 }
    \caption{Our version of the TAAL Trojan can replace a logic locking key with a malign key that allows the leakage of secret data. Unlikely input combinations trigger the Trojan. The logic locking key bit is replaced with a 0 in this example.\label{fig:taal_trojan}\vspace{-0.4cm}}
\end{figure}
\fi

The adversary's strategy can be divided into two stages: Analysis and attack. First, they analyze the method for propagating encryption keys to the primary outputs. Next, using the gathered information, they attack the activated circuit.

\paragraph{\textbf{Analysis}}
For the first phase, we assume the adversary has access to the logic-locked netlist. The netlist can be retrieved either by directly accessing the design (external design houses and foundries) or by reverse engineering the chip. Access to the IC can be remote or physical. In addition, we assume the adversary...
\begin{itemize}
    \item ... knows the location of logic-locking input ports.
    \item ... knows the position of all signals carrying sensitive information, such as encryption keys.
    \item ... does not know the design's exact functionality, as it cannot be extracted from the logic-locked design.
\item ... can observe the outputs of the design remotely or with direct access to the chip.
\end{itemize}
In Fig.~\ref{fig:path_sensitization}, an adversary obtained access to the logic-locked netlist and leverages an ATPG framework to generate a pattern, namely $lolo\_key1 = 1$ and $lolo\_key2 = 1$. This particular pattern facilitates the propagation of the secret $enc\_key1$ to the observable output $output2$. Notably, upon analysis, it is evident that $enc\_key2$ is not susceptible to leakage for any input pattern. Subsequently, the attacker stores the generated patterns for use in the second phase of the attack.

The adversary is able to manipulate the logic-locking key inputs by tampering with the storage holding the LL activation keys or modifying the value before it reaches the key gates. \textcolor{red}{}It is often assumed that a tamper-proof memory is utilized to protect the activation key. However, hardware Trojans, including TAAL~\cite{taal}, may exploit vulnerabilities near the key storage to leak the key after activation. 

Additionally, attacks that leverage fault injection to alter the contents of the logic-locking storage are also possible.

\paragraph{\textbf{Attack}}
The adversary has access to an activated manufactured chip \textcolor{red}{}as an end-user. Using the manipulation techniques, the attacker can change the key in the activated IC for a short time to gain access to secret data, such as encryption keys or user data. In this work, we prove that this access to sensitive data is introduced by LL. Then, the original functionality is restored by applying the correct LL key. 

\section{Related Work}
To the best of our knowledge, \textit{this work is the first to evaluate the influence of LL on the confidentiality property in secure hardware designs using an automated methodology}. Recently, logic locking has been exploited to break the integrity of a neural accelerator during runtime. Logic locking is used as a backdoor in this context to reduce the quantity of the correct detections~\cite{lolo_ai_attack}. A manual inspection of the logic-locked MiG-V processor revealed that incorrect LL keys can be used to leak sensitive data like encryption keys~\cite{mig_v_cracked}.

Additionally, the application of path sensitization used in this work has been utilized in other contexts of LL. For example, path sensitization has been applied to analyze whether LL key bits are stored safely~\cite{path_sensitization_lolo}. 
Furthermore, SAT-attacks (a popular class of key-retrieval attacks) aim to solve Boolean satisfiability problems to assess how inputs propagate from primary inputs to primary outputs~\cite{sat_attack1, sat_attack2, sat_attack3, sat_on_lolo}. Retrieving the LL key bits can unlock the locked netlist and enable IP piracy, overproduction, and hardware Trojans.
\textit{Nevertheless, a comprehensive analysis of the impact of LL on the security properties of the unlocked design is still missing.} We address this research gap by developing an automated methodology to evaluate this impact and showcase it for three representative LL algorithms. 
\section{Methodology}
\label{ch:methodology}
\textcolor{red}{}This study is investigating the effect of logic locking on the confidentiality property of hardware. To achieve this, we analyze cryptographic designs---circuits explicitly engineered to uphold this property and often used in logic locking research due to their complex dataflow. These serve as baseline for evaluating the impact of logic locking on ICs with less stringent security measures. Cryptographic keys are treated as the secret in this work, a concept applicable to other areas like filters (taps) and neural networks (weights). OpenCores \cite{opencores} offers DES, GOST, XTEA, and KECCAK-32 implementations. An AES-128 Verilog design is provided by the platform Trust-Hub \cite{trusthub}. Logic-locking schemes use randomization combined with a set of rules to place the key-driven logic. Thus, a set of obfuscated benchmarks needs to be generated to allow a suitable statistical analysis of the occurrence of vulnerabilities.
The evaluation 
can be separated into the following steps:\\
\myCircled{1} Use ATPG to find the secret bits that can be read at the output using stuck-at-fault tests on non-locked benchmarks.\\
\myCircled{2}Apply the three state-of-the-art LL schemes on the chosen benchmarks. Generate a set of locked benchmarks allowing a statistical analysis for each benchmark and algorithm.\\
\myCircled{3}Perform step \myCircled{1} on the set of obfuscated benchmarks.\\
\myCircled{4}Compare results for the locked and non-locked designs. \\
\myCircled{5}Evaluate the vulnerabilities manually for each leakage.\\
\myCircled{6}Compare the security of different LL techniques.

An adversary would only conduct step \myCircled{3} on the single obfuscated IC. Step \myCircled{2} is further elaborated below. 

\subsection{Preparing the Benchmarks}
The evaluation considers five Verilog hardware designs that implement cryptographic algorithms. While AES, DES, GOST, and XTEA represent encryption algorithms, KECCAK-32 implements a hashing method.
\begin{table}[b!]
    \centering
    \vspace{-0.3cm}
        \caption{Information about the signals that are labeled secret for the underlying benchmarks.\label{tab:enc_key_sizes}}
    \begin{tabular}{c|c|c | c|c| c}
          & \multicolumn{5}{c}{Benchmarks} \\
          & AES & DES & GOST & KECCAK & XTEA  \\\hline\hline
        Size& 128 bits & 56 bits & 256 bits &  32 bits &  128 bits\\\hline
        Type & Enc. key & Enc. key & Enc. key & Input & Enc. key  \\

        \end{tabular}
\end{table}
\begin{table}[b!]
    \centering
    \vspace{-0.3cm}
        \caption{ASSURE's logic locking key sizes for all benchmarks and encryption modes. \label{tab:assure_lolo_key_size}}
    \begin{tabular}{c|c|c | c|c| c}
        ASSURE & \multicolumn{5}{c}{Benchmarks (logic locking key sizes in bits)} \\
        locking modes & AES & DES & GOST & KECCAK & XTEA  \\\hline\hline
        Branch & - & 768 & 1 & 12 & 2 \\\hline
        Ops & 373 & 33 & 2 & 95 & 61 \\\hline
        Const &704512 & 32768& 517&184 & 684  \\\hline
        Branch+Ops & -& 801 & 3& 107 & 63 \\\hline
        Branch+Const &-& 33536& 518& 196 & 686  \\\hline
        Const+Ops &704885& 32801 & 519 & 279 & 745  \\\hline
        All & - & 33569 & 520 & 291 & 747 \\
        \end{tabular}
   
\end{table}
As shown in Table~\ref{tab:enc_key_sizes}, for the encryption methods, the encryption key is labeled as the secret signal that is considered in this work. The different key sizes used by the algorithms are listed as well. For the hashing algorithm, the input data are labeled as sensitive signals.
Now, each benchmark is obfuscated using the three logic-locking algorithms: ASSURE, EPIC, and D-MUX (see Section~\ref{sec-ll-desc}). The resulting logic-locking key lengths are explained  below.

\paragraph{\textbf{ASSURE}} Each of the three locking mechanisms (constant locking, operation locking, and branch locking) is evaluated individually. All combinations of the mechanisms are evaluated (constants + operations, branches + operations, branches + constants, all three). As there is a limited amount of branches, operations, and constants in a design, the number of locking locations is limited. For the LL, all possible placement locations are used, resulting in the LL key sizes (see Table~\ref{tab:assure_lolo_key_size}). 

\paragraph{\textbf{EPIC}} 
The RTL benchmarks are synthesized into gate-level netlists. These non-logic-locked benchmarks are used for the first evaluation. 
Furthermore, the netlists locked with EPIC can be grouped according to the key length, with 100\% representing the maximum number of key placement locations for the benchmark. However, gate-level locking techniques can use a considerably higher number of gate insertion points than ASSURE. 
Relative key sizes of 1\%, 25\%, and 50\% are chosen. These key sizes reflect the standard overhead assumed in LL evaluations.
\begin{table}[b!]
    \centering
    \vspace{-0.4cm}
        \caption{EPIC's logic locking key sizes for all benchmarks and encryption modes. \label{tab:epic_lolo_key_size}}
    \begin{tabular}{c|c|c | c|c| c}
        Relative & \multicolumn{5}{c}{Benchmarks (logic locking key sizes in bits)} \\
        key size & AES & DES & GOST & KECCAK & XTEA  \\\hline\hline
        1\% & 4177 & 523  & 25 &242  & 100 \\\hline
       25\% & 104415 & 13064 & 631 & 6043 & 2500 \\\hline
        50\% & 208830 & 26127 & 1262 & 12086 & 4999  \\
        \end{tabular}
\end{table}

Compared to the ASSURE evaluation, not all possible key gate placements are used, which results in numerous possibilities to lock the same original gate level netlist.
Simply assessing one netlist cannot provide a guarantee against the possibility of leakage created by the LL method. 
Therefore, we analyze test sets of 1,000 locked netlists to ensure the comprehensive coverage of potential vulnerabilities.

\paragraph{\textbf{D-MUX}}
The handling process is similar to that of EPIC. For a thorough evaluation, we generate 1,000 individually logic-locked benchmarks for each relative key size. However, the process of inserting multiplexers with D-MUX is complex as every inserted multiplexer must not create combinational cycles in the design. Therefore, generating D-MUX-locked benchmarks with large key sizes is not viable\footnote{Note that this is a limitation of the D-MUX scheme, not our evaluation.}. Consequently, smaller relative key sizes are chosen for evaluation (0.5\% and 1\%). The absolute key sizes selected for evaluation are listed in Table~\ref{tab:dmux_lolo_key_size}.
\begin{table}[b!]
    \centering
    \vspace{-0.3cm}
        \caption{D-MUX's logic locking key sizes for all benchmarks and encryption modes. \label{tab:dmux_lolo_key_size}}
    \begin{tabular}{c|c|c|c|c| c}
        Relative & \multicolumn{5}{c}{Benchmarks (logic locking key sizes in bits)} \\
        key size & AES & DES & GOST & KECCAK & XTEA  \\\hline\hline
        0.5\% & 1503 & 184 & 9 & 99 & 37  \\\hline
        1\% & 3006 & 268 & 17 & 198 & 75 \\
        \end{tabular}
\end{table}
The aforementioned detection procedure to identify possible data leakages is explained below.
\subsection{Confidentiality Attack: Detecting Leakages}
\label{sec:det_leakages}
We use TestMAX~\cite{testmax} to analyze the possibility of creating input patterns that would sensitize the secret encryption keys or hashing inputs to the primary outputs of the hardware. We define two attack scenarios:
\begin{itemize}
\item \textsc{Set-All}: The attacker can set all the inputs of the IC, except the secret bits.
\item \textsc{Set-Ll-Key}: The attacker can modify the LL key to forward the sensitive data in the circuit. 
\end{itemize}
\begin{figure}[t!]
    \centering
    \includegraphics[width=\columnwidth]{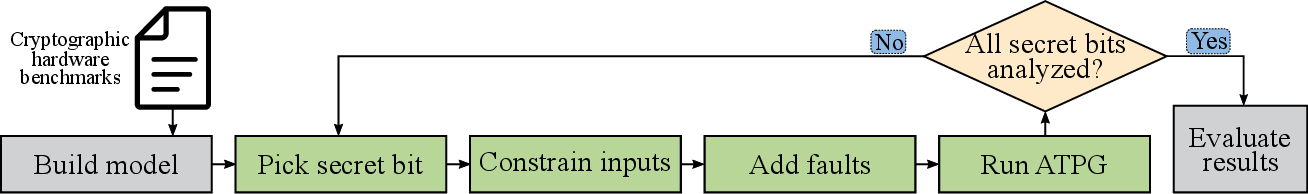 }
    \caption{The ATPG framework is used to identify leakage paths for each sensitive bit. \label{fig:evaluation_procedure}}
\vspace{-0.3cm}

\end{figure}
\begin{figure*}

\begin{subfigure}{0.49\textwidth}
\centering
\begin{tikzpicture}[scale=1]
    \begin{axis}[
        ybar stacked,
        enlargelimits=false, 
        scaled y ticks = false,
        bar width=0.045cm,
        width=9.2cm,
        height=3.8cm,
        style={align=center}, ylabel=Leakage distribution (\%),
        xlabel = Secret bits,
        ymax=100,
        ymin=-0.050,
        enlarge x limits=0.02,
        xtick={0,16,32,48,64,80,96,112,128},
        ylabel near ticks,
        ylabel shift={-0.55em},
        ytick pos=left,
        y tick label style={font=\small},
        y label style={font=\small, xshift=-0.75em}
    ]
    \addplot[fill=rwth_red!50, draw=none,] table {data/ATPG_mux_1_XTEA_timeout_10_sec_set_all_bits_dt.txt};
    \addplot[fill=rwth_blue!50, draw=none,] table {data/ATPG_mux_1_XTEA_timeout_10_sec_set_all_bits_nd.txt};
    \addplot[fill=rwth_green!50, draw=none,] table {data/ATPG_mux_1_XTEA_timeout_10_sec_set_all_bits_au.txt};

    \legend{DT, ND, S}
    \end{axis}
    \end{tikzpicture}  
    \vspace{-0.55cm}
    \caption{\label{fig:hist_bitwise_dmux_1_set_all}\textsc{Set\_All} histogram for D-MUX 1\%.\vspace{0.1cm}}
\end{subfigure}
\begin{subfigure}{0.49\textwidth}
\centering
\begin{tikzpicture}[scale=1]
    \begin{axis}[
        ybar stacked,
        enlargelimits=false, 
        scaled y ticks = false,
        bar width=0.045cm,
        width=9.2cm,
        height=3.8cm,
        style={align=center}, ylabel=Leakage distribution (\%),
        xlabel = Secret bits,
        ymax=100,
        ymin=-0.050,
        enlarge x limits=0.02,
        xtick={0,16,32,48,64,80,96,112,128},
        ylabel near ticks,
        ylabel shift={-0.55em},
        ytick pos=left,
        y tick label style={font=\small},
        y label style={font=\small, xshift=-0.75em}
    ]
    \addplot[fill=rwth_red!50, draw=none,] table {data/ATPG_mux_1_XTEA_timeout_10_sec_set_ll_bits_dt.txt};
    \addplot[fill=rwth_blue!50,draw=none,] table {data/ATPG_mux_1_XTEA_timeout_10_sec_set_ll_bits_nd.txt};
    \addplot[fill=rwth_green!50, draw=none,] table {data/ATPG_mux_1_XTEA_timeout_10_sec_set_ll_bits_au.txt};

    \legend{DT, ND, S}
    \end{axis}
    \end{tikzpicture}  
        \vspace{-0.55cm}

    \caption{\label{fig:hist_bitwise_dmux_1_set_ll}\textsc{Set-Ll-Key} histogram for D-MUX 1\%.\vspace{0.1cm}}
\end{subfigure}

\begin{subfigure}{0.49\textwidth}
\centering
\begin{tikzpicture}[scale=1]
    \begin{axis}[
        ybar stacked,
        enlargelimits=false, 
        scaled y ticks = false,
        bar width=0.045cm,
        width=9.2cm,
        height=3.8cm,
        style={align=center}, ylabel=Leakage distribution (\%),
        xlabel = Secret bits,
        ymax=100,
        ymin=-0.050,
        enlarge x limits=0.02,
        xtick={0,16,32,48,64,80,96,112,128},
        ylabel near ticks,
        ylabel shift={-0.55em},
        ytick pos=left,
        y tick label style={font=\small},
        y label style={font=\small, xshift=-0.75em}
        ]
    \addplot[fill=rwth_red!50, draw=none,] table {data/ATPG_xor-xnor_25_XTEA_timeout_12_sec_set_all_bits_dt.txt};
    \addplot[fill=rwth_blue!50, draw=none,] table {data/ATPG_xor-xnor_25_XTEA_timeout_12_sec_set_all_bits_nd.txt};
    \addplot[fill=rwth_green!50, draw=none,] table {data/ATPG_xor-xnor_25_XTEA_timeout_12_sec_set_all_bits_au.txt};

    \legend{DT, ND, S}
    \end{axis}
    \end{tikzpicture}  
        \vspace{-0.55cm}

    \caption{\label{fig:hist_bitwise_epic_25_set_all}\textsc{Set\_All} histogram for EPIC 25\%.}
\end{subfigure}
\begin{subfigure}{0.49\textwidth}
\centering
\begin{tikzpicture}[scale=1]
    \begin{axis}[
        ybar stacked,
        enlargelimits=false, 
        scaled y ticks = false,
        bar width=0.045cm,
        width=9.2cm,
        height=3.8cm,
        style={align=center}, ylabel=Leakage distribution (\%),
        xlabel = Secret bits,
        ymax=100,
        ymin=-0.050,
        enlarge x limits=0.02,
        xtick={0,16,32,48,64,80,96,112,128},
        ylabel near ticks,
        ylabel shift={-0.55em},
        ytick pos=left,
        y tick label style={font=\small},
        y label style={font=\small, xshift=-0.75em}
    ]
    \addplot[fill=rwth_red!50, draw=none,] table {data/ATPG_xor-xnor_25_XTEA_timeout_12_sec_set_ll_bits_dt.txt};
    \addplot[fill=rwth_blue!50, draw=none,] table {data/ATPG_xor-xnor_25_XTEA_timeout_12_sec_set_ll_bits_nd.txt};
    \addplot[fill=rwth_green!50, draw=none,] table {data/ATPG_xor-xnor_25_XTEA_timeout_12_sec_set_ll_bits_au.txt};

    \legend{DT, ND, S}
    \end{axis}
    \end{tikzpicture}  
        \vspace{-0.55cm}

    \caption{\label{fig:hist_bitwise_epic_25_set_ll}\textsc{Set-Ll-Key} histogram for EPIC 25\%.}
\end{subfigure}

\caption{\label{fig:hist_bitwise}The histogram illustrates the leakage distribution for each secret bit in the 1,000 generated benchmarks for one key size of each EPIC and D-MUX. (Not detected [ND], Secure [S], and Detected [DT]). DT indicates that the confidential bit can be transferred to an output. Bits marked as S are not transferable, and no pattern has been identified yet for ND tests.}
\vspace{-0.4cm}

\end{figure*}
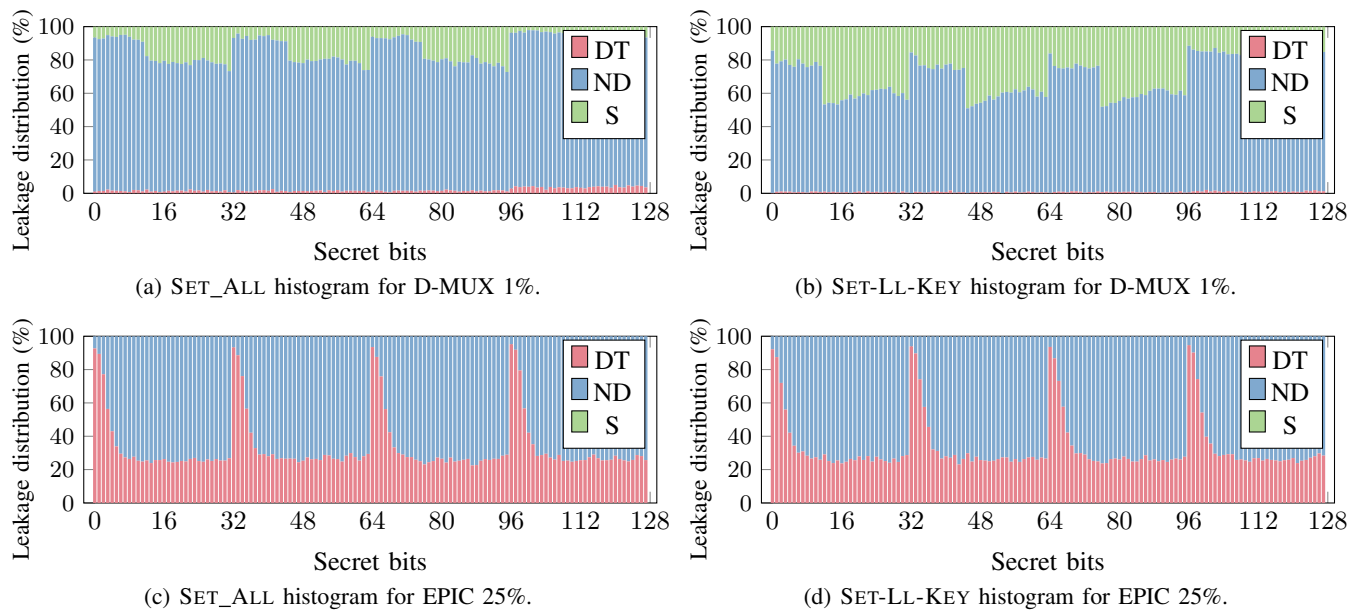
The evaluation process for each bit of the secret is described in the following steps and illustrated in Fig.~\ref{fig:evaluation_procedure}. \\
    \myCircled{1} \emph{Build model:} The gate-level netlist file and technology library are loaded to build the model in the ATPG framework. 
    \myCircled{2} \emph{Pick secret bit:} Decide which signal is analyzed. Label it.\\
    \myCircled{3} \emph{Constrain inputs:} In order to simulate the proposed attack model, inputs that are not accessible to the adversary must be constrained with the unknown value ("X"). The first attack scenario involves constraining only the secret inputs of the benchmark, while the second scenario sets the inputs of the benchmark to the unknown value, leaving only the logic-locking key inputs available for pattern generation.\\
    \myCircled{4} \emph{Label secrets:} The secret bit is marked.\\
    \myCircled{5} \emph{Run ATPG:} To propagate the secret bit to the primary outputs using unconstrained inputs, the ATPG framework generates input patterns. \\
    \myCircled{6} \emph{Repetition}: Steps  \myCircled{2} -  \myCircled{5} are repeated for each bit.\\
    \myCircled{6} \emph{Evaluate the results:} After ATPG is performed, the bits are assigned to one of the following classes: \textit{Detected (DT), Secure (S), and Not Detected (ND)}.

As the framework aims to detect vulnerabilities rather than verify the security of each gate-level netlist, only hard-detected (DT) tests are considered security threats. Although this simplification does not guarantee that all ND tests are secure and do not create vulnerabilities, using the same test setup for all benchmarks allows for comparison. Furthermore, identifying the minimum number of vulnerabilities allows a first threat evaluation, which is the primary goal of the evaluation.

\section{Evaluation}
First, we conduct an analysis of the five benchmarks without LL, with the aim of determining if any confidential information can be unintentionally transmitted to the primary outputs of the designs. The investigation revealed that all secret bits were secure, \textit{implying that the original benchmarks do not expose any sensitive data to potential adversaries.}
Moving forward, we examine the sets of logic-locked benchmarks and compare the results with the leak-proof nature of the non-logic-locked benchmarks. This allows for pinpointing any vulnerabilities that may have been introduced by the LL algorithms.

\subsection{Leakage Distribution Analysis}
Cryptographic algorithms are sensitive to the inputs they receive. This includes encryption key bits and hashing inputs. Similarly, the leakage caused by LL varies depending on the location of the secret bit. To highlight this point, we have depicted the leakage distribution for the XTEA cryptographic benchmark of each bit of the 128-bit encryption key in a set of 1,000 logic-locked designs. The analysis was conducted for every key size and algorithm combination of D-MUX and EPIC, and a selection of the results can be seen in Fig.~\ref{fig:hist_bitwise}.
Since there is only one benchmark available for each ASSURE mode, no bitwise analysis is presented. The leakages introduced by ASSURE will be discussed in later sections.

The evaluations are conducted on a cluster of computers using Ryzen 3900X processors. Each test is given a 10-second analysis time. Although the maximum runtime of the ATPG per test was increased to 13 hours for some of the tests, the test remained ND (see Fig.~\ref{fig:hist_bitwise}). Therefore, the overall maximum runtime was not adapted as such an analysis has to be conducted for a total of approx. 6,000,000 tests. While the ND bits may still be at risk of being leaked, they are not treated as a leakage in this work since a more detailed examination would be too laborious. Therefore, the presented leakage rates may be even higher for the algorithm and key size combination than what is shown in this work.
Also, as shown by the red sub bars, \textit{both EPIC and D-MUX introduced leakages into the hardware designs for both attack scenarios}, \textsc{Set-All} and \textsc{Set-Ll-Key}. By modifying the inputs, secret bits can be leaked to the design's outputs. While only LL key bits can be modified to attack the secret in \textsc{Set-Ll-Key}, secret information can be leaked to accessible outputs for the benchmarks for both algorithms.

\begin{figure*}
\begin{center}
\includegraphics[width=0.7\columnwidth]{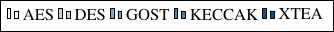}
\vspace{-0,25cm}
\end{center}
\begin{subfigure}{0.49\textwidth}
\centering
\begin{tikzpicture}[scale=1]
    \begin{axis}[
        ybar,
        enlargelimits=false, 
        scaled y ticks = false,
        legend style={at={(0.495,1.3)}, anchor=north,legend columns=-1, font=\footnotesize},
        bar width=0.18cm,
        width=9cm,
        height=3.5cm,
        style={align=center}, ylabel=Average\\detection rate(\%),
        ymax=2.800,
        ymin=-0.0,
        enlarge x limits=0.43,
        symbolic x coords={\textsc{0.5\%},\textsc{1\%}},
        xtick=data,
        ylabel near ticks,
        ylabel shift={-0.55em},
        xtick pos=left,
        ytick pos=left,
        legend image post style={scale=0.6},
        y tick label style={font=\small},
        y label style={font=\small}
    ]
    \addplot[fill=white] coordinates {(\textsc{0.5\%},0.0078125) (\textsc{1\%},0.0140625)};
    \addplot[fill=rwth_blue!25] coordinates {(\textsc{0.5\%},0.06607143) (\textsc{1\%},0.2125)};
    \addplot[fill=rwth_blue!50] coordinates {(\textsc{0.5\%},0.80390625) (\textsc{1\%},1.741015625)};
    \addplot[fill=rwth_blue!75] coordinates {(\textsc{0.5\%},1.409375) (\textsc{1\%},2.66875)};
    \addplot[fill=rwth_blue] coordinates {(\textsc{0.5\%},1.03203125) (\textsc{1\%},1.98671875)};

    \end{axis}
    \end{tikzpicture}  
    \vspace{-0.2cm}
    \caption{\label{fig:aver_dmux_set_all}\textsc{Set-All} average for D-MUX\vspace{0.25cm}}
\end{subfigure}
\begin{subfigure}{0.49\textwidth}
\centering
\begin{tikzpicture}[scale=1]
    \begin{axis}[
        ybar,
        enlargelimits=false, 
        scaled y ticks = false,
        legend style={at={(0.495,1.3)}, anchor=north,legend columns=-1, font=\footnotesize},
        bar width=0.18cm,
        width=9cm,
        height=3.5cm,
        style={align=center}, ylabel=Average\\detection rate(\%),
        ymax=0.9,
        ymin=-0.0050,
        enlarge x limits=0.43,
        symbolic x coords={\textsc{0.5\%},\textsc{1\%}},
        xtick=data,
        ylabel near ticks,
        ylabel shift={-0.55em},
        xtick pos=left,
        ytick pos=left,
        legend image post style={scale=0.6},
        y tick label style={font=\small},
        y label style={font=\small}
    ]
    \addplot[fill=white] coordinates {(\textsc{0.5\%},0.00078125) (\textsc{1\%},0.01074129)};
    \addplot[fill=rwth_blue!25] coordinates {(\textsc{0.5\%},0.04285714) (\textsc{1\%},0.1)};
    \addplot[fill=rwth_blue!50] coordinates {(\textsc{0.5\%},0.00976563) (\textsc{1\%},0.02539063)};
    \addplot[fill=rwth_blue!75] coordinates {(\textsc{0.5\%},0.0) (\textsc{1\%},0.0125)};
    \addplot[fill=rwth_blue] coordinates {(\textsc{0.5\%},0.42578125) (\textsc{1\%},0.82890625)};

    \end{axis}
    \end{tikzpicture}  
        \vspace{-0.2cm}

    \caption{\label{fig:aver_dmux_set_ll}\textsc{Set-Ll-Key} average for D-MUX\vspace{0.25cm}}
\end{subfigure}

\begin{subfigure}{0.49\textwidth}
\centering
\begin{tikzpicture}[scale=1]
    \begin{axis}[
        ybar,
        enlargelimits=false, 
        scaled y ticks = false,
        legend style={at={(0.495,1.3)}, anchor=north,legend columns=-1, font=\footnotesize},
        bar width=0.18cm,
        width=9cm,
        height=3.5cm,
        style={align=center}, ylabel=Average\\detection rate(\%),
        ymax=89.000,
        ymin=-0.050,
        enlarge x limits=0.13,
        symbolic x coords={\textsc{1\%},\textsc{25\%}, \textsc{50\%}},
        xtick=data,
        ylabel near ticks,
        ylabel shift={-0.55em},
        xtick pos=left,
        ytick pos=left,
        legend image post style={scale=0.6},
        y tick label style={font=\small},
        y label style={font=\small}
    ]
    \addplot[fill=white] coordinates {(\textsc{1\%},0.0) (\textsc{25\%},0) (\textsc{50\%},0.0)};
    \addplot[fill=rwth_blue!25] coordinates {(\textsc{1\%},0.0) (\textsc{25\%},69.67) (\textsc{50\%},52.94)};
    \addplot[fill=rwth_blue!50] coordinates {(\textsc{1\%},0.4257) (\textsc{25\%},86.57) (\textsc{50\%},81.74)};
    \addplot[fill=rwth_blue!75] coordinates {(\textsc{1\%},1.084) (\textsc{25\%},42.08) (\textsc{50\%},19.75)};
    \addplot[fill=rwth_blue] coordinates {(\textsc{1\%},2.70) (\textsc{25\%},33.75) (\textsc{50\%},43.74)};

    \end{axis}
    \end{tikzpicture}  
       \vspace{-0.2cm}
 
    \caption{\label{fig:aver_epic_set_all}\textsc{Set-All} Average for EPIC\vspace{0.25cm}}
\end{subfigure}
\begin{subfigure}{0.49\textwidth}
\centering
\begin{tikzpicture}[scale=1]
    \begin{axis}[
        ybar,
        enlargelimits=false, 
        scaled y ticks = false,
        legend style={at={(0.495,1.3)}, anchor=north,legend columns=-1, font=\footnotesize},
        bar width=0.18cm,
        width=9cm,
        height=3.5cm,
        style={align=center}, ylabel=Average\\detection rate(\%),
        ymax=50.00,
        ymin=-0.050,
        enlarge x limits=0.13,
        symbolic x coords={\textsc{1\%},\textsc{25\%}, \textsc{50\%}},
        xtick=data,
        ylabel near ticks,
        ylabel shift={-0.55em},
        xtick pos=left,
        ytick pos=left,
        legend image post style={scale=0.6},
        y tick label style={font=\small},
        y label style={font=\small}
    ]
    \addplot[fill=white] coordinates {(\textsc{1\%},0.0) (\textsc{25\%},0.0) (\textsc{50\%},0.0)};
    \addplot[fill=rwth_blue!25] coordinates {(\textsc{1\%},0.0) (\textsc{25\%},0.0) (\textsc{50\%},0.0)};
    \addplot[fill=rwth_blue!50] coordinates {(\textsc{1\%},0.0) (\textsc{25\%},4.833) (\textsc{50\%},20.0)};
    \addplot[fill=rwth_blue!75] coordinates {(\textsc{1\%},0.0) (\textsc{25\%},0.80) (\textsc{50\%},4.49)};
    \addplot[fill=rwth_blue] coordinates {(\textsc{1\%},0.9625) (\textsc{25\%},33.79) (\textsc{50\%},43.86)};

    \end{axis}
    \end{tikzpicture}  
        \vspace{-0.2cm}

    \caption{\label{fig:aver_epic_set_ll}\textsc{Set-LL-Key} Average for EPIC\vspace{0.25cm}}
\end{subfigure}

\begin{subfigure}{\textwidth}
\centering
\begin{tikzpicture}[scale=1]
    \begin{axis}[
        ybar,
        enlargelimits=false, 
        scaled y ticks = false,
        legend style={at={(0.495,1.3)}, anchor=north,legend columns=-1},
        bar width=0.18cm,
        width=18cm,
        height=3.5cm,
        style={align=center}, ylabel=Average\\detection rate(\%),
        ymax=105.000,
        ymin=-0.050,
        enlarge x limits=0.05,
        symbolic x coords={\textsc{Branch},\textsc{Ops}, \textsc{Consts}, \textsc{Branch+Ops}, \textsc{Branch+Consts}, \textsc{Consts+Ops}, \textsc{All}},
        xtick=data,
        ylabel near ticks,
        ylabel shift={-0.55em},
        xtick pos=left,
        ytick pos=left,
        legend image post style={scale=0.9},
        y tick label style={font=\small},
        y label style={font=\small}
    ]
    \addplot[fill=white] coordinates {(\textsc{Branch},0.0) (\textsc{Ops},0) (\textsc{Consts},0.0) (\textsc{Branch+Ops},0.0) (\textsc{Branch+Consts},0.0) (\textsc{Consts+Ops},25.5625) (\textsc{All},0.0)};
    \addplot[fill=rwth_blue!25] coordinates {(\textsc{Branch},0.0) (\textsc{Ops},0) (\textsc{Consts},0.0) (\textsc{Branch+Ops},0.0) (\textsc{Branch+Consts},0.0) (\textsc{Consts+Ops},0.0) (\textsc{All},0.0)};
    \addplot[fill=rwth_blue!50] coordinates {(\textsc{Branch},0.0) (\textsc{Ops},0) (\textsc{Consts},0.0) (\textsc{Branch+Ops},0.0) (\textsc{Branch+Consts},0.0) (\textsc{Consts+Ops},0.0) (\textsc{All},0.0)};
    \addplot[fill=rwth_blue!75] coordinates {(\textsc{Branch},0.0) (\textsc{Ops},43.75) (\textsc{Consts},100.0) (\textsc{Branch+Ops},28.125) (\textsc{Branch+Consts},0.0) (\textsc{Consts+Ops},100.0) (\textsc{All},56.25)};
    \addplot[fill=rwth_blue] coordinates {(\textsc{Branch},0.0) (\textsc{Ops},0.78125) (\textsc{Consts},10.15) (\textsc{Branch+Ops},1.5625) (\textsc{Branch+Consts},10.15) (\textsc{Consts+Ops},64.06) (\textsc{All},60.94)};

    \end{axis}
    \end{tikzpicture}  
    
    \caption{\label{fig:aver_assure_set_all}\textsc{Set-All} Average for ASSURE. No detection for \textsc{Set-Ll-Key}}
\end{subfigure}
\caption{\label{fig:average_detections} Relative average detection rate for all 3 LL algorithms, for multiple key sizes and 5 benchmarks. \vspace{-0.4cm}}
\end{figure*}
Moreover, the examination of Fig.s~\ref{fig:hist_bitwise_dmux_1_set_all} and \ref{fig:hist_bitwise_dmux_1_set_ll} demonstrates that D-MUX is more prone to leak the later bits of the XTEA encryption key. However, EPIC indicates that the first bits of a 32-bit segment in the 128-bit secret are more susceptible to the XOR-XNOR locking scheme. Comparing Fig.~\ref{fig:hist_bitwise_dmux_1_set_all} with Fig.~\ref{fig:hist_bitwise_dmux_1_set_ll} shows that the number of leakages is diminished for the \textsc{Set-Ll-Key} attack scenario relative to the \textsc{Set-All}, as expected. A higher number of inputs that can be modified to leak information leads to a greater success rate. 

Overall, each secret key bit can be forwarded to the output of the design with a probability of at least 20\% for the EPIC algorithm (25\% relative key size). However, this does not imply that all bits are vulnerable in the same benchmark of the set. For the smaller key sizes of the D-MUX algorithm, leakages in roughly 5\% of all benchmarks can be achieved for the later bits.
The following evaluation only focuses on the detections (DT) indicating a leakage. The AU and ND tests are both labeled as non-leakages.

\subsection{Average Detection Rate Analysis}
The resulting average detection rates for all benchmarks are depicted in Fig.~\ref{fig:average_detections}.
For the graphs, the number of average detections in the set is divided by the length of the secret, allowing a comparison between the benchmarks. The effect of LL varies with the benchmark.

High average detection rates can be observed for the EPIC algorithm (see Fig.~\ref{fig:aver_epic_set_all} and Fig.~\ref{fig:aver_epic_set_ll}), which can use a bigger relative key size. Area-optimized designs like the XTEA and GOST benchmark are impacted significantly by the EPIC algorithm. The designs utilize FSMs that control the computation before releasing the ciphertext to the output. When XOR or XNOR gates are placed by EPIC in this control engine, the key gates can be used to flip bits and misuse some of the computational steps to allow a leakage of the data before it has been obfuscated sufficiently. 
Although the average detection rate seems to be reduced for the higher relative key size of 50\% compared to the 25\% one (see Fig.~\ref{fig:aver_epic_set_all}), it does not mean that fewer vulnerabilities are present in the benchmarks with more key gates. As explained before, the ND bits are not elaborated here. The ND tests require a longer runtime to decide whether the bits can be leaked (DT) or cannot be forwarded to the outputs of the design (S). The averages reflect the number of leakages that are at least present in the logic-locked benchmarks.
\begin{figure*}
\begin{subfigure}{0.49\textwidth}
\centering
\begin{tikzpicture}[scale=1]
    \begin{axis}[
        ybar,
        enlargelimits=false, 
        scaled y ticks = false,
        bar width=0.055cm,
        width=9.2cm,
        height=3.5cm,
        style={align=center}, ylabel=\# of benchmarks,
        xlabel = Number of bits leaked in a design,
        ymax=300,
        ymin=-0.050,
        enlarge x limits=0.02,
        xtick={0,16,32,48,64,80,96,112,128},
        ylabel near ticks,
        ylabel shift={-0.55em},
        ytick pos=left,
        y tick label style={font=\small},
        y label style={font=\small, xshift=-0.5em},
        extra x ticks={0,46},
        extra x tick labels={},
        extra x tick style={
            xmajorgrids=true,
            xtick style={
                /pgfplots/major tick length=0pt,
            },
            grid style={
                red,
                /pgfplots/on layer=axis foreground,
            },
            }
    ]
    \addplot[fill=rwth_blue!75, draw=none,] table {data/ATPG_mux_1_XTEA_timeout_10_sec_set_all_distr.txt};

    \end{axis}
    \end{tikzpicture}  
        \vspace{-0.55cm}
    \caption{\label{fig:histogram_dmux_1_set_all}\textsc{Set\_All} histogram for D-MUX 1\%}
\end{subfigure}
\begin{subfigure}{0.49\textwidth}
\centering
\begin{tikzpicture}[scale=1]
    \begin{axis}[
        ybar,
        enlargelimits=false, 
        scaled y ticks = false,
        bar width=0.055cm,
        width=9.2cm,
        height=3.5cm,
        style={align=center}, ylabel=\# of benchmarks,
        xlabel = Number of bits leaked in a design,
        ymax=450,
        ymin=-0.050,
        enlarge x limits=0.02,
        xtick={0,16,32,48,64,80,96,112,128},
        ylabel near ticks,
        ylabel shift={-0.55em},
        ytick pos=left,
        y tick label style={font=\small},
        y label style={font=\small, xshift=-0.5em},
        extra x ticks={0,24},
        extra x tick labels={},
        extra x tick style={
            xmajorgrids=true,
            xtick style={
                /pgfplots/major tick length=0pt,
            },
            grid style={
                red,
                /pgfplots/on layer=axis foreground,
            },
            }
    ]
    \addplot[fill=rwth_blue!75, draw=none,] table {data/ATPG_mux_1_XTEA_timeout_10_sec_set_ll_distr.txt};

    \end{axis}
    \end{tikzpicture}  
        \vspace{-0.55cm}
    \caption{\label{fig:histogram_dmux_1_set_ll}\textsc{Set-Ll-Key} histogram for D-MUX 1\%}
\end{subfigure}

\begin{subfigure}{0.49\textwidth}
\centering
\begin{tikzpicture}[scale=1]
    \begin{axis}[
        ybar,
        enlargelimits=false, 
        scaled y ticks = false,
        bar width=0.055cm,
        width=9.2cm,
        height=3.5cm,
        style={align=center}, ylabel=\# of benchmarks,
        xlabel = Number of bits leaked in a design,
        ymax=85,
        ymin=-0.050,
        enlarge x limits=0.02,
        xtick={0,16,32,48,64,80,96,112,128},
        ylabel near ticks,
        ylabel shift={-0.55em},
        ytick pos=left,
        y tick label style={font=\small},
        y label style={font=\small, xshift=-0.5em},
        extra x ticks={23,75},
        extra x tick labels={},
        extra x tick style={
            xmajorgrids=true,
            xtick style={
                /pgfplots/major tick length=0pt,
            },
            grid style={
                red,
                /pgfplots/on layer=axis foreground,
            },
            }
    ]
    \addplot[fill=rwth_blue!75, draw=none] table {data/ATPG_xor-xnor_50_XTEA_timeout_10_sec_set_all_distr.txt};

    \end{axis}
    \end{tikzpicture}  
        \vspace{-0.25cm}
    \caption{\label{fig:histogram_epic_50_set_all}\textsc{Set\_All} histogram for EPIC 50\%}
\end{subfigure}
\begin{subfigure}{0.49\textwidth}
\centering
\begin{tikzpicture}[scale=1]
    \begin{axis}[
        ybar,
        enlargelimits=false, 
        scaled y ticks = false,
        bar width=0.055cm,
        width=9.2cm,
        height=3.5cm,
        style={align=center}, ylabel=\# of benchmarks,
        xlabel = Number of bits leaked in a design,
        ymax=85,
        ymin=-0.050,
        enlarge x limits=0.02,
        xtick={0,16,32,48,64,80,96,112,128},
        ylabel near ticks,
        ylabel shift={-0.55em},
        ytick pos=left,
        y tick label style={font=\small},
        y label style={font=\small, xshift=-0.5em},
        extra x ticks={24,78},
        extra x tick labels={},
        extra x tick style={
            xmajorgrids=true,
            xtick style={
                /pgfplots/major tick length=0pt,
            },
            grid style={
                red,
                /pgfplots/on layer=axis foreground,
            },
            }
    ]
    \addplot[fill=rwth_blue!75, draw=none] table {data/ATPG_xor-xnor_50_XTEA_timeout_12_sec_set_ll_distr.txt};

    \end{axis}
    \end{tikzpicture}  
        \vspace{-0.25cm}
    \caption{\label{fig:histogram_epic_50_set_ll}\textsc{Set-Ll-Key} histogram for EPIC 50\%}
\end{subfigure}
\caption{\label{fig:histogram} Histogram illustrating the distribution of the netlists over the number of leakages for the XTEA benchmark. The red lines indicate the netlist(s) with the lowest highest number of leakages for the combination of key size and locking scheme. }
\vspace{-0.3cm}
\end{figure*}
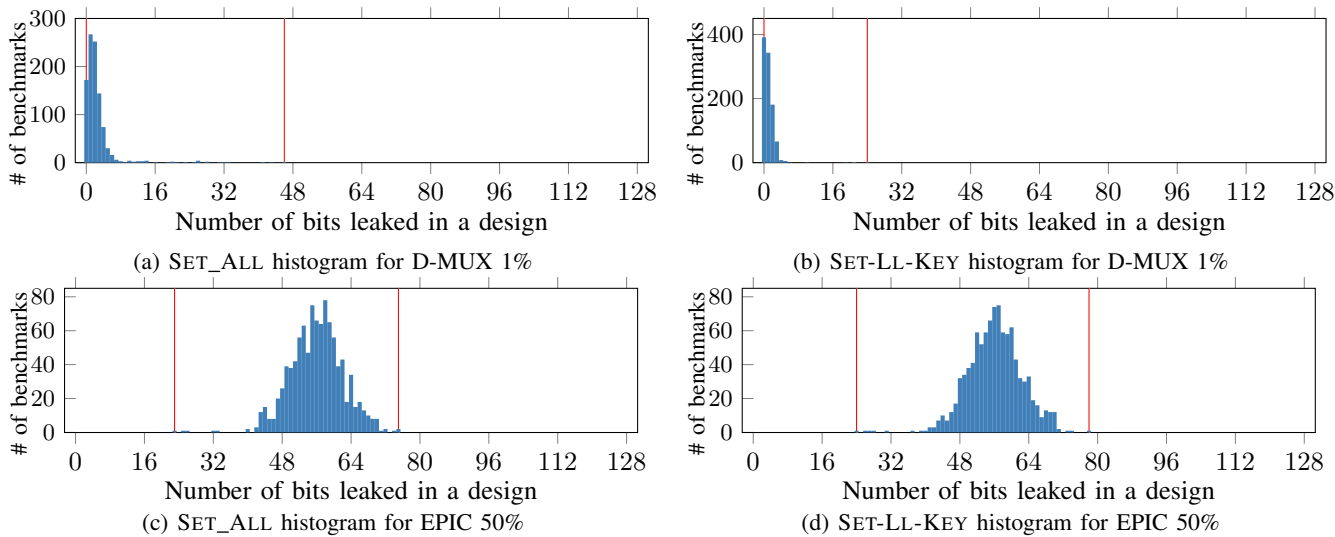
ASSURE shows the highest relative leakage in the KECCAK benchmark in most locking modes for the \textsc{Set-All} attack scenario, as illustrated in Fig.~\ref{fig:aver_assure_set_all}. Three of the five crypto benchmarks show significant leakages introduced by ASSURE (KECCAK, XTEA and AES), while DES and GOST remain secure. This shows the impact of the hardware design on the introduction of vulnerabilities using LL. 
The average detection rate for the EPIC 1\% benchmarks is lower than 10\% for all cryptographic algorithms. However, only considering the average is not sufficient, and outliers must be considered for a comprehensive threat assessment. Therefore, a histogram analysis is conducted to illustrate the number of leakages in each logic-locked gate-level netlist.
\subsection{Histogram Analysis}
To provide a clearer understanding of the number of leakages in each logic-locked netlist, Fig.~\ref{fig:histogram} presents a histogram analysis of the XTEA benchmark, which enables a comprehensive comparison of the algorithms across all relative key sizes and attack scenarios. Regrettably, no set of netlists is available for the ASSURE implementation, so no histogram analysis could be conducted for it. 
Despite the fact that the complete set yields an average lower than 10\%, a closer look at the distributions, depicted in Fig.~\ref{fig:histogram_dmux_1_set_all} and Fig.~\ref{fig:histogram_dmux_1_set_ll}, reveals that a significant portion of the encryption key can still be compromised even for a smaller number of relative key sizes. 
\textit{While the average detection rate provides an initial assessment of the security risk posed by LL schemes, it is worth noting that outliers highlight the true extent of their impact on the confidentiality property of an IC, which can be catastrophic.}
The evaluation of EPIC with higher relative key sizes finds that the minimum number of leakages exceeds 20 bits out of the 128-bit key. This implies that regardless of the key gates' placement in XTEA among the 1,000 benchmarks, modifying the LL key would leak at least 20 bits (Fig.~\ref{fig:histogram_epic_50_set_ll}).

\subsection{Limitations and Future Work}
 As mentioned before, a relatively high number of ND tests are still present in the final results. The evaluation for removing the ND tests takes a significant amount of time. No feasible increase of the time limit for the ATPG framework resulted in the change of the tests' label. \textit{However, it means that the number of introduced leakages into the benchmark can be even higher than presented here.} Quantitative information flow analysis methods may identify additional vulnerabilities~\cite{qflow, qflow2}. These frameworks use probabilistic analysis to quantify the amount of information an adversary can gain about a secret by observing the outputs.


Possible mitigations include iterative approaches that analyze the security properties after LL the circuit are a possibility. If vulnerabilities are identified, the circuit would need to be logic-locked again. 

\section{Conclusion}

In this study, we investigated the impact of logic locking on the confidentiality of sensitive signals in hardware descriptions. \textit{Through path sensitization, we found that some cryptographic benchmarks, which were deemed secure before the application of logic locking, exhibited major data leakages after logic locking was applied}. While ASSURE is relatively secure against attacks that modify only the LL key, it can still leak up to 100\% of the key when all inputs are under the attacker's control. Compared to ASSURE, EPIC exhibits a significantly higher susceptibility to leakage, with up to 73.83\% of the encryption key being compromised solely by modifying the logic locking key. Furthermore, D-MUX leaks only up to 25\% of the encryption key in the same attack scenario. \textit{Therefore, it is evident that logic locking can pose a significant risk to the confidentiality of sensitive data in hardware designs.} 
Nonetheless, we acknowledge logic locking's ability to protect the IC from hardware Trojans throughout the supply chain, albeit at the cost of compromising confidentiality.

\bibliographystyle{IEEEtran}
\bibliography{bibtexentry}

\end{document}